\documentclass[a4paper,10pt]{article}
\usepackage[a4paper, margin=2cm]{geometry}

\usepackage[utf8]{inputenc}
\usepackage[T1]{fontenc}

\usepackage{fourier}

\usepackage{amsmath}
\allowdisplaybreaks
\usepackage{amsfonts}
\usepackage{amssymb}
\usepackage{stmaryrd}
\usepackage{mathrsfs}
\usepackage{bm}
\usepackage{booktabs}
\usepackage{graphicx}

\usepackage[hidelinks]{hyperref}

\usepackage{amsthm}
\theoremstyle{remark}
\newtheorem*{remark}{Remark}

\makeatletter
\def\th@remark{%
	\thm@headfont{\bfseries}%
	\normalfont 
	\thm@preskip\topsep \divide\thm@preskip\tw@
	\thm@postskip\thm@preskip
}
\makeatother

\graphicspath{{Figures/}}

\makeatletter
\newcommand*{\transpose}{%
	{\mathpalette\@transpose{}}%
}
\newcommand*{\@transpose}[2]{%
	\raisebox{\depth}{$\m@th#1\intercal$}%
}
\makeatother

\usepackage[authoryear,round]{natbib}
\usepackage{doi}
\usepackage{url}

\title{On the accuracy of one-way approximate models\\ for nonlinear waves in soft solids}

\author{Harold Berjamin \textsuperscript{a} \\
	{\footnotesize
	\textsuperscript{a}School of Mathematical and Statistical Sciences, University of Galway, University Road, Galway, Republic of Ireland}
}

\date{}

\begin{document}
	
\maketitle{}

\begin{abstract}
	
	\noindent
	Simple strain-rate viscoelasticity models of isotropic soft solid are introduced. The constitutive equations account for finite strain, incompressibility, material frame-indifference, nonlinear elasticity, and viscous dissipation. A nonlinear viscous wave equation for the shear strain is obtained exactly, and corresponding one-way Burgers-type equations are derived by making standard approximations. Analysis of the travelling wave solutions shows that these partial differential equations produce distinct solutions, and that deviations are exacerbated when wave amplitudes are not arbitrarily small. In the elastic limit, the one-way approximate wave equation can be linked to simple wave theory and shock wave theory, thus allowing direct error measurements.
	
\end{abstract}


\section{Introduction}\label{sec:Intro}

In nonlinear acoustics, the Burgers equation is often viewed as the simplest model equation that includes nonlinear wave propagation and diffusion effects \citep{whitham99}. This partial differential equation in space and time can be derived directly from the one-dimensional Navier--Stokes equation by dropping the pressure term, or as a special case of the Westervelt equation. Besides Burgers' equation, other one-way wave equations have been derived to describe wave propagation in fluids and solids at large amplitudes \citep{hamilton98,naugolnykh98}. Based on an appropriate scaling of the wave amplitude, such approximate partial differential equations describe unidirectional wave motion for slowly-varying wave profiles of moderate amplitude.

One-way approximate wave equations have found applications in various areas of nonlinear acoustics. For instance, works by \citet{radostin13} and \citet{nazarov17} describe compression wave propagation in solids with bimodular elastic behaviour. Another example is the Zabolotskaya equation that describes unidirectional plane shear wave propagation in soft solids such as gels and brain tissue \citep{zabo04}, see also \citet{cormack18}. In these latter cases, the underlying three-dimensional constitutive theories were revisited by \citet{destrade13} as well as \citet{saccomandi21} to enforce objectivity (i.e., invariance by change of observer), leading to slight modifications of the equations of motion.

For these partial differential equations, not many analytical solutions are known. Nevertheless, it is sometimes possible to derive exact stationary wave solutions that keep an invariant wave profile throughout the motion, which occurs at a suitable constant speed. Those permanent waveforms result from the interaction between nonlinearity and dispersion (here of dissipative nature), a common feature that they share with solitary waves.

One might wonder whether it is preferable to seek closed-form travelling wave solutions by using directly the full equations of motion, or by using their one-way approximation. As a matter of fact, both approaches have been considered separately in the above literature. The present study aims to provide evidence to advocate for a derivation of travelling waves based on the complete equations of motion, thus supporting a remark by \citet{jordan05a} in relation with the study by \citet{catheline03}{\,---\,}this remark led to the publication of an erratum that briefly discusses the validity of a particular one-way wave equation \citep{catheline03e}.\footnote{The wave equation proposed by \citet{catheline03} and analysed by \citet{jordan05a} cannot be obtained rigorously from the equations of motion unless spatial derivatives $\partial/\partial z$ are replaced by $-\frac1{c}\partial/\partial t$, a relationship that is only valid for travelling waves of speed $c$.}

For this purpose, we consider the case of shear wave propagation in soft viscoelastic solids of strain rate type. We derive the simplest isotropic constitutive theories that account for finite strain, incompressibility, material frame-indifference, and viscous dissipation (Section~\ref{sec:Constitutive}). Then, this framework is applied to simple shear deformations, aka. transverse plane waves (Section~\ref{sec:Shear}), including the reduction to a one-way model described by a Burgers-type equation with cubic nonlinearity. Finally, we investigate the travelling wave solutions deduced from the full equations of motion as well as from the reduced wave equations (Section~\ref{sec:Travelling}). Results show non-negligible discrepancies introduced by the reduction to unidirectional motion as soon as wave amplitudes are no longer infinitesimal. These comparisons are reconsidered in the lossless elastic limit where connections between the one-way model and other theories are established (Section~\ref{sec:Elast}).

\section{Strain-rate model}\label{sec:Constitutive}

\subsection{Basic equations}\label{subsec:Basic}

In what follows, we present the basic equations of Lagrangian dynamics for incompressible solids \citep{holzapfel00}. We consider a homogeneous and isotropic solid continuum on which no external body force is applied. Its motion in the Euclidean space is described by using an orthonormal Cartesian coordinate system $(O,x,y,z)$. Thus, a particle initially located at some position $\bm{X}$ of the reference configuration moves to a position $\bm{x}$ of the current configuration. The deformation gradient is the second-order tensor defined as $\bm{F} = {\partial \bm{x}}/{\partial \bm{X}}$. Introducing the displacement field $\bm{u} = \bm{x} - \bm{X}$ and the identity tensor $\bm{I} = [\delta_{ij}]$ whose components are represented by the Kronecker delta function, the relationship $\bm{F} = \bm{I} + \nabla \bm{u}$ is obtained, where $\nabla$ denotes the gradient operator with respect to the material coordinates ${\bm{X}} = (x,y,z)$.

For \emph{incompressible solids}, volume does not change during deformation, which is expressed by the constraint
\begin{equation}
	J = \det\bm{F} \equiv 1 .
	\label{J}
\end{equation}
Thus, the mass density $\rho$ is constant in time. It follows also that $\dot{J} = J\,\text{tr}\,\bm{L} \equiv 0$, where the dot denotes the material time derivative $\partial/\partial t$. The notation $\text{tr} \,\bm{L}$ stands for the trace $L_{ii}$ of the Eulerian velocity gradient tensor $\bm{L} = [L_{ij}]$, where summation over repeated indices was assumed (Einstein notation). The components $L_{ij} = \dot{F}_{ik}{F}^{-1}_{kj}$ of $\bm{L} = \dot{\bm F}\bm{F}^{-1}$ are obtained by matrix multiplication.

Various strain tensors are defined as functions of $\bm F$. Here, constitutive laws are expressed in terms of the Green--Lagrange strain tensor $\bm{E} = \tfrac12 (\bm{F}^{\transpose} \bm{F} - \bm{I})$, where $\bm{F}^{\transpose} = [F_{ji}]$ denotes the transpose of $\bm F$. Indeed, the finite strain tensor
\begin{equation}
	\bm{E} = \tfrac12 (\nabla\bm{u}+\nabla^\transpose\bm{u}+\nabla^\transpose\bm{u}\, \nabla\bm{u})
\end{equation}
is often a preferred choice in physical acoustics, see for instance \citet{zabo04}. We introduce also its rate $\dot{\bm E} = \bm{F}^{\transpose} \bm{D} \bm{F}$ obtained by differentiation with respect to time, where $\bm{D} = \tfrac12 (\bm{L} + \bm{L}^{\transpose})$ is the strain rate tensor. We note that $\bm{D}$ is trace-free due to incompressibility \eqref{J}.

The motion is governed by the conservation of linear momentum equation $\rho \dot{\bm v} = \nabla\cdot \bm{P}$, where $\bm v = \dot{\bm x}$ is the velocity field and $\rho$ is the mass density. The equation of motion involves the Lagrangian divergence of the first Piola--Kirchhoff stress tensor $\bm{P} = \bm{F}\bm{S}$ where $\bm{S} = \bm{S}^\transpose$ is the second Piola--Kirchhoff stress tensor. Those stress tensors are specified later on by the provision of a constitutive law.

The present definitions are consistent with notations and conventions used in the monograph by \citet{holzapfel00}. In particular, the divergence of the tensor $\bm P$ reads $[\nabla\cdot \bm{P}]_i = P_{ij,j}$ componentwise, where indices after the comma denote spatial differentiation. In some other texts, a transposed definition of the divergence is used. Then, the equation of motion involves the material divergence of the nominal stress tensor $\bm{P}^\transpose$ instead of $\bm P$.

\subsection{Generalities}\label{subsec:General}

In the present study, we consider isotropic deformable solids whose constitutive behaviour is described by the state variables ${\mathfrak S} = \lbrace s, \bm{E}\rbrace$, where $s$ is the specific entropy. The choice of variables ${\mathfrak S}$ is coherent with the postulate of frame-indifference of the internal energy \citep{holzapfel00}. In fact, a change of observer specified by a superimposed rigid-body motion leaves ${\mathfrak S}$ invariant, as well as the internal energy $U$. Note that the internal energy does not depend on rates of strain.

The internal energy per unit volume $U$ is a function of state to be specified. The thermodynamic temperature is defined as the conjugate variable of $s$ in the partial Legendre transform of $U/\rho$ with respect to $s$ \citep{berjamin20}. However, the explicit dependence of $U$ with respect to $s$ is usually omitted in the definition of a strain energy density function $W^\text{e}$ such that $U = W^\text{e}(\bm{E})$.
The strain energy $W^\text{e}$ is regarded as a scalar-valued \emph{isotropic function} of its arguments. Thus, its dependence with respect to $\bm{E}$ can be reduced to a dependence with respect to three scalar invariants
\begin{equation}
	I_1 = \text{tr}(\bm{E}), \quad I_2 = \text{tr}(\bm{E}^2), \quad I_3 = \text{tr}(\bm{E}^3) .
	\label{InvarElast}
\end{equation}
They can be used directly, or other physically meaningful scalar quantities might be defined from them.

The first and second principles of thermodynamics yield the Clausius--Duhem inequality
\begin{equation}
	\mathscr{D} = \text{tr}\big( (\bm{S}-\bm{S}^\text{e}) \dot{\bm E} \big) = \text{tr}\big( \bm{S}^\text{v} \dot{\bm E} \big) \geq 0 ,
	\label{CD}
\end{equation}
where $\mathscr{D}$ is the dissipation, $\bm{S} = \bm{S}^\text{e} + \bm{S}^\text{v}$ is the total second Piola--Kirchhoff stress,
\begin{equation}
	\bm{S}^\text{e} = -p\bm{C}^{-1} + \frac{\partial W^\text{e}}{\partial \bm{E}}
	\label{Constitutive}
\end{equation}
denotes the elastic part, and $\bm{S}^\text{v}$ is a viscous contribution to be specified subsequently. The scalar $p$ is an arbitrary Lagrange multiplier for the incompressibility constraint \eqref{J}, see Sec.~6.3 of \citet{holzapfel00}, and $\bm{C}=\bm{I}+2\bm{E}$ is the right Cauchy--Green strain tensor $\bm{F}^{\transpose} \bm{F}$ (i.e., its inverse is given by $\bm{C}^{-1} = \bm{F}^{-1}\bm{F}^{-\transpose}$). Therefore, according to \eqref{CD}, no dissipation occurs in the elastic case $\bm{S}=\bm{S}^\text{e}$ where the viscous stress tensor $\bm{S}^\text{v}$ is equal to zero.

According to the dissipation inequality \eqref{CD}, the viscous stress $\bm{S}^\text{v}$ is a function of state and evolution variables, e.g. the set ${\mathfrak S} \cup \lbrace \dot{\bm E}\rbrace$ which is a consistent choice to enforce frame-indifference \citep{ball02,destrade13}. We introduce a \emph{dissipation potential} $W^\text{v}(\bm{E}, \dot{\bm E})$ such that
\begin{equation}
	\bm{S}^\text{v} = \frac{\partial W^\text{v}}{\partial \dot{\bm E}}
	\label{ConstitutiveViscous}
\end{equation} defines the viscous stress \citep{maugin99}.
In general, the dissipation potential is described by additional invariants \citep{pioletti00}
\begin{equation}
	\begin{aligned}
		& I_4 = \text{tr}(\dot{\bm E}), \quad I_5 = \text{tr}(\dot{\bm E}^2), \quad I_6 = \text{tr}(\dot{\bm E}^3),\\
		& I_7 = \text{tr}(\dot{\bm E}\bm{E}), \quad I_8 = \text{tr}(\dot{\bm E}\bm{E}^2), \quad I_9 = \text{tr}(\dot{\bm E}^2\bm{E}),\\ & I_{10} = \text{tr}(\dot{\bm E}^2\bm{E}^2) .
	\end{aligned}
	\label{InvarVisco}
\end{equation}
In the present study, we consider Newtonian-type viscosity models whose dissipation potential is as simple as possible.

\subsection{Consequences of incompressibility}

First, let us investigate the consequences of the incompressibility constraint \eqref{J}. As noted in \citet{jacob07}, the invariants \eqref{InvarElast} of $\bm{E}$ are linked through
\begin{equation}
	I_1 = I_2 - \tfrac43 I_3 - I_1^2 + 2I_1I_2 - \tfrac23 I_1^3 \, ,
	\label{InvarElastIncomp}
\end{equation}
by virtue of incompressibility. This identity follows from the expression of the principal invariants of the unimodular tensor $\bm{C} = \bm{I} + 2\bm{E}$ in terms of the invariants $I_k$, see the Appendix of \citet{destrade10c}. Using the differential version of the incompressibility constraint, the invariants \eqref{InvarElast}-\eqref{InvarVisco} of $\bm{E}$, $\dot{\bm E}$ satisfy the particular relationship
\begin{equation}
	\tfrac12 I_4 = I_{7} + 2I_1I_{7} - 2 I_{8} - (I_1 - I_2 + I_1^2) I_4
	\label{InvarViscoIncomp}
\end{equation}
deduced from the identity $\text{tr}\, \bm{D} = 0$, see Appendix.

The relationship \eqref{InvarElastIncomp} means that the invariant $I_1 = \text{tr}(\bm{E})$ is no longer linear with respect to the components of the strain tensor $\bm E$; instead, Eq.~\eqref{InvarElastIncomp} shows that it has terms of polynomial order two and three with respect to the strain. Furthermore, due to the relationship \eqref{InvarViscoIncomp}, the invariant $I_4 = \text{tr}(\dot{\bm E})$ is still linear with respect to the components of the strain-rate tensor $\dot{\bm E}$. However, Eq.~\eqref{InvarViscoIncomp} shows that $I_4$ is no longer invariant on the strain tensor $\bm E$; instead, it has terms of polynomial order one, two and three with respect to the Green--Lagrange strain.

\subsection{Constitutive assumptions}

In weakly nonlinear elasticity, the strain energy density function is sought in the form of a polynomial of the invariants $I_k$ with constant coefficients. Similarly to \citet{zabo04}, we assume that the internal energy $U$ has a fourth-order polynomial expression with respect to the components of the strain tensor $\bm E$ of the form
\begin{equation}
	W^\text{e} = \mu I_2 + \tfrac13 A I_3 + D I_2^2 ,
	\label{StrainEnergy}
\end{equation}
where $\mu\geq 0$ is the shear modulus (in Pa), and the coefficients $A$, $D$ are higher-order elastic constants. Note in passing that this approach can be extended to the modelling of transversely isotropic elastic solids, see discussions in \citet{destrade10b}.

Now, let us propose an expression for the dissipation potential. To end up with viscosity models similar to that by \citet{destrade13}, we assume that the dissipation potential is a second-order polynomial expansion of the strain rate tensor $\dot{\bm E}$, and a zeroth-order polynomial of $\bm E$. This assumption amounts to selecting $W^\text{v}$ of second order in $({\bm E},\dot{\bm E})$, and to ignore the terms proportional to $\dot{\bm E}$ that produce elastic stresses. Due to the relationships \eqref{InvarElastIncomp}-\eqref{InvarViscoIncomp}, we therefore keep
\begin{equation}
	W^\text{v} = \eta I_5  ,
	\label{DissipationPotential}
\end{equation}
where $\eta \geq 0$ is the shear viscosity (in Pa.s). In the above expression, the absence of bulk viscosity ``$\zeta$'' is due to the assumption on polynomial orders for the viscous part, and to the incompressibility property \eqref{InvarViscoIncomp}. Setting the bulk viscosity $\zeta = \frac23 \eta$ in \citet{destrade13} yields the same expressions as above.

Computation of the tensor derivatives of the potentials \eqref{StrainEnergy}-\eqref{DissipationPotential} by means of the chain rule for $W^{\bullet}(I_k, \dots)$ yields the following elastic \eqref{Constitutive} and viscous stress contributions \eqref{ConstitutiveViscous}
\begin{equation}
	\begin{aligned}
		\bm{S}^\text{e} &= -p \bm{C}^{-1} + 2 (\mu + 2 D I_2) \bm{E} + A \bm{E}^2 ,\\
		\bm{S}^\text{v} &= 2 \eta \dot{\bm E} .
	\end{aligned}
	\label{Stresses}
\end{equation}
Thermodynamic consistency \eqref{CD} is ensured provided that the dissipation $\mathscr{D} = 2W^\text{v}$ is non-negative. In fact, the present dissipation potential $W^\text{v}$ is a homogeneous function of degree two with respect to $\dot{\bm E}$ \citep{maugin99}. A sufficient condition for the restriction $\mathscr{D} \geq 0$ to be always satisfied is that the viscosity $\eta$ is non-negative.

In complement, we consider the alternative dissipation potential $W^\text{v} = \eta\, \text{tr}(\bm{D}^{2})$ which is related to the mechanics of Newtonian fluids. This quantity is equal to $\eta\, \text{tr}(\dot{\bm \Pi}\dot{\bm E})$, where we have introduced the rate of a Piola-type strain tensor $\bm{\Pi} = \bm{C}^{-1}\bm{E}$ (see Appendix). The viscous stress \eqref{ConstitutiveViscous} takes the Newtonian form
\begin{equation}
	\bm{S}^\text{v} = 2\eta \dot{\bm \Pi},
	\label{Fluid}
\end{equation}
more commonly written $\bm{T}^\text{v} = 2\eta\bm{D}$ in terms of the Cauchy stress tensor $\bm{T}^\bullet = \bm{F}\bm{S}^\bullet\bm{F}^\transpose$. This expression is also perfectly satisfactory from the point of view of material frame-indifference \citep{destrade13}. By using similar arguments as above, thermodynamic consistency \eqref{CD} is ensured if the viscosity $\eta$ is non-negative. For further details, the interested reader is referred to \citet{antman88}.

In the infinitesimal strain limit, the equations of motion are linearised with respect to the components of $\bm{u}$ and $\nabla\bm{u}$, which are assumed arbitrarily small. The same linearisation procedure is also applied to the constitutive equations $\bm{P} = \bm{P}^\text{e}+\bm{P}^\text{v}$ with $\bm{P}^\bullet = \bm{F}\bm{S}^\bullet$. Using the previous constitutive laws, we end up with the linearised expressions $\bm{P}^\text{e} \simeq -p\bm{I} + 2\mu \bm{\varepsilon}$ and $\bm{P}^\text{v} \simeq 2\eta\dot{\bm\varepsilon}$ of the elastic and viscous stresses, where $\bm{\varepsilon} = \frac12 (\nabla\bm{u}+\nabla^\transpose\bm{u})$ is the infinitesimal strain tensor.

\section{Plane shear waves}\label{sec:Shear}

\subsection{Nonlinear viscous wave equation}\label{subsec:ShearWave}

Similarly to \citet{destrade13}, we consider simple shear deformations described by the displacement field $\bm{u} = [u,0,0]^\transpose$ where $u=u(z,t)$ denotes the particle displacement along the $x$-direction. Thus, the deformation gradient tensor reads
\begin{equation}
	\bm{F} = [\delta_{ij} + u_{i,j}] =
	{
		\begin{bmatrix}
			1 & 0 & \gamma\\
			0 & 1 & 0 \\
			0 & 0 & 1
	\end{bmatrix}} ,
\end{equation}
where $\gamma = \partial u/\partial z$ is the shear strain. The velocity field takes the form $\bm{v} = [v,0,0]^\transpose$ where $v = \partial u/\partial t$ is the shear velocity.

In the equation of motion $\rho \dot{\bm v} = \nabla\cdot \bm{P}$, the relevant first Piola--Kirchhoff stress component $P_{13}$ is deduced from the elastic part $P_{13}^\text{e} = \mu \gamma + \Gamma \gamma^3$ where $\Gamma = \mu + A/2 + D$ is a parameter of nonlinearity, and from the viscous part. Keeping only terms up to order $\gamma^3$, the viscous stress in \eqref{Stresses}\textsubscript{b} yields $P_{13}^\text{v} = \eta (1+2\gamma^2) \dot\gamma$, whereas the Newtonian stress \eqref{Fluid} yields $P_{13}^\text{v} = \eta \dot\gamma$. Thus, to combine both models in one expression, we will use $P_{13}^\text{v} = \eta (1+\delta\gamma^2) \dot\gamma$ where $\delta \in \lbrace 0,2\rbrace$ is a parameter.

Finally, upon division by the shear modulus $\mu$, the $x$-component of the equation of motion produces the nonlinear wave equation
\begin{equation}
	\begin{aligned}
		\frac{1}{c^2} \frac{\partial^2 u}{\partial t^2} &= \frac{\partial^2 u}{\partial z^2} + \frac23 \beta \frac{\partial}{\partial z} \left(\frac{\partial u}{\partial z}\right)^3 \\
		&\; + \tau \frac{\partial}{\partial z} \left[ \left(1 + \delta\left(\frac{\partial u}{\partial z}\right)^2\right) \frac{\partial^2 u}{\partial z\partial t}\right] ,
	\end{aligned}
	\label{Wave}
\end{equation}
describing transverse wave propagation along the $z$-direction, where we have introduced the notations
\begin{equation}
	c = \sqrt{\frac{\mu}{\rho}}, \quad \beta = \frac32 \frac{\Gamma}{\mu}, \quad \tau = \frac{\eta}{\mu} .
	\label{WaveCoeffs}
\end{equation}
Spatial differentiation of Eq.~\eqref{Wave} allows to write a similar wave equation for the strain
\begin{equation}
	\frac{1}{c^2} \frac{\partial^2 \gamma}{\partial t^2} = \frac{\partial^2 \gamma}{\partial z^2} + \frac23\beta \frac{\partial^2}{\partial z^2} \gamma^3 + \tau \frac{\partial^2}{\partial z^2} \left[ \left(1 + \delta\gamma^2\right) \frac{\partial \gamma}{\partial t}\right]
	\label{WaveStrain}
\end{equation}
Here, we have obtained the same wave equations than those derived in \citet{destrade13} for the particular bulk viscosity $\zeta = \frac23 \eta$. Note in passing the presence of a nonlinear viscous term when $\delta = 2$.

According to the wave equation \eqref{Wave}, the propagation of infinitesimal shear waves is governed by the linear Kelvin--Voigt wave equation $c^{-2}u_{tt} = u_{zz}+\tau u_{zzt}$. Dispersion analysis shows that the complex velocity of a linear harmonic wave equals $c \sqrt{1+\text{i}\omega\tau}$ where $\omega$ is the angular frequency and $\text{i}$ is the imaginary unit. Thus, in the low-frequency range, such a wave propagates at the shear wave speed $c$ in the absence of nonlinearity. Typically, this sound velocity equals $c \approx 2$~m/s in gels \citep{jacob07}, whereas $\beta \approx 10$ and $\tau \approx 0.12$~ms at a loading frequency of $100$~Hz.

For smooth initial- and boundary-value problems, the existence of global smooth solutions to \eqref{Wave} with $\delta = 0$ is established. However, it is worth pointing out that the picture is radically different in the case $\delta = 2$ for which such a solution might not exist for arbitrarily large times \cite{maccamy70}. This property is particularly inconvenient for the study of long-time quasi-static solutions, thus indicating that \eqref{Fluid} is the more appropriate form of the viscous stress. \citet{pucci11} illustrate the blow-up of smooth solutions in finite time for $\delta=2$ and $\beta=0$, and further links to the mathematical literature are provided therein.

\subsection{Slow scale approximations}\label{subsec:Slow}

Similarly to \citet{zabo04} and \citet{pucci19}, we proceed now to a reduction of the above wave equation \eqref{Wave} for one-way wave propagation with slowly varying profile. We present two approximations based either on a slow space variable or a slow time variable.

\paragraph{Slow space}

Let us follow the scaling procedure in \citet{zabo04}. For this purpose, we introduce the following \emph{scaling} defined by the change of variables $\lbrace \tilde z = \epsilon^2 z, \tilde t = t-z/c , u = \epsilon\tilde u\rbrace$, where $\epsilon$ is a small parameter and $\tilde u = \tilde u(\tilde z, \tilde t)$. Furthermore, we assume that $\tau$ is of order $\epsilon^2$. Note that this set of assumptions corresponds to a slowly-varying profile in space.

This Ansatz is then substituted in the equation of motion \eqref{Wave}. At leading (cubic) order in $\epsilon$, the motion of soft viscous solids is governed by the scalar equation
\begin{equation}
	\epsilon^3 c \frac{\partial^2 \tilde u}{\partial \tilde z\partial \tilde t} = \epsilon^3\frac{\beta}{c^2} \left(\frac{\partial \tilde u}{\partial \tilde t}\right)^2 \frac{\partial^2 \tilde u}{\partial \tilde t^2} + \epsilon\frac{\tau}{2} \frac{\partial^3 \tilde u}{\partial \tilde t^3}.
\end{equation}
Transforming back to the initial displacement $u$ and physical coordinates $(z,t)$ leads to a reduced wave equation
\begin{equation}
	c\frac{\partial v}{\partial z} + \left(1 - \beta v^2/c^2\right) \frac{\partial v}{\partial t} = \frac{\tau}{2} \frac{\partial^2 v}{\partial t^2} ,
	\label{Burgers}
\end{equation}
for the velocity $v = \partial  u/\partial t$.

Up to the choice of time variable used here (i.e., the physical time $t$ instead of the retarded time $\tilde t$), the partial differential equation \eqref{Burgers} is identical to the cubic Burgers-type equation of \citet{zabo04}. However, the extra nonlinear viscosity term $\delta\tau\, \partial (\gamma^2 \dot \gamma)/\partial z$ is not apparent in the one-way approximation \eqref{Burgers}. In fact, this additional term is lost in the rescaling procedure given that it is of higher order in $\epsilon$ than the leading-order viscous term $\tau\, \partial \dot \gamma/\partial z$. In the end, setting $\delta=2$ in the wave equation \eqref{Wave} does not induce any modification of the transport equation \eqref{Burgers}.

\paragraph{Slow time}

For later comparisons, let us derive a similar Burgers-type equation governing the evolution of the strain instead of the velocity by following \citet{pucci19}. To do so, we introduce the slow-time scaling based on the change of variables $\lbrace \tilde t = \epsilon^2 t, \tilde z = z-ct , u = \epsilon\tilde u\rbrace$ where $\epsilon$ is a small parameter. Proceeding in a similar fashion to above, we end up with the nonlinear transport equation
\begin{equation}
	\frac{\partial \gamma}{\partial t} + c \left(1 + \beta \gamma^2\right) \frac{\partial \gamma}{\partial z} = \frac{\tau c^2}{2} \frac{\partial^2 \gamma}{\partial z^2} ,
	\label{SlowStrain}
\end{equation}
where $\gamma = \partial u/\partial z$ is the shear strain. Here too, after keeping leading order terms, we have transformed back to the initial physical coordinates $(z,t)$. Therefore, the above partial differential equation may be viewed as a one-way approximation of the wave equation \eqref{WaveStrain}. Their travelling wave solutions are compared in the next section; the inviscid limit $\tau=0$ is discussed in Sec.~\ref{sec:Elast}.

\section{Travelling wave solutions}\label{sec:Travelling}

\begin{figure}
	\centering
	\includegraphics{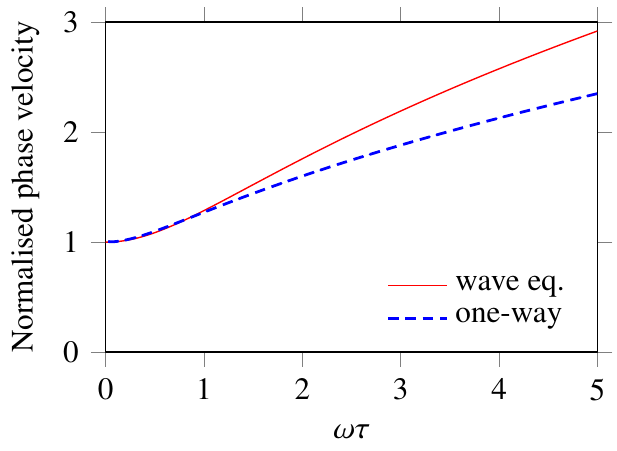}
	
	\caption{Infinitesimal strain limit. Evolution of the phase velocity with the dimensionless frequency $\omega\tau$. \label{fig:Dispersion}}
\end{figure}

Let us begin with a brief remark on wave dispersion \citep{carcione15}. According to the slow time approximation \eqref{SlowStrain}, the propagation of infinitesimal shear waves is governed by the advection-diffusion equation $\gamma_t + c \gamma_z = \frac{\tau}2 c^2 \gamma_{zz}$. It follows that the complex velocity of a harmonic wave reads $\frac{c}2 (1 + \sqrt{1 + 2 \text{i}\omega \tau})$. Comparison with the viscous wave equation of Sec.~\ref{subsec:ShearWave} shows that the one-way approximation is accurate in the low-frequency range only, where it provides a correct estimation of the wave velocity up to order $O(\omega\tau)$\,---\,the same analysis for Eq.~\eqref{Burgers} leads to a similar conclusion. This property is illustrated in Fig.~\ref{fig:Dispersion} where the evolution of the phase velocities divided by $c$ is displayed. With this observation in mind, we focus now on the study of nonlinear solutions.

\subsection{Nonlinear viscous wave equation}\label{subsec:KinkWave}

Let us seek travelling wave solutions to the wave equation \eqref{WaveStrain}, i.e. specific smooth waveforms that propagate at a constant velocity with a steady profile. In a similar fashion to \citet{destrade13}, we first introduce the following rescaled dimensionless variables and coordinates
\begin{equation}
	g(\bar z,\bar t) = \sqrt{\tfrac23\beta}\,\gamma(z, t), \quad \bar t = t/\tau, \quad \bar z = z/(c\tau),
	\label{Scale}
\end{equation}
in Eq.~\eqref{WaveStrain}, such that
\begin{equation}
	\frac{\partial^2 g}{\partial \bar t^2} = \frac{\partial^2 g}{\partial \bar z^2} + \frac{\partial^2}{\partial \bar z^2} g^3 +  \frac{\partial^2}{\partial \bar z^2} \left[ \left(1 + \frac{3\delta}{2\beta} g^2\right) \frac{\partial g}{\partial \bar t}\right] .
	\label{WaveStrainScaled}
\end{equation}
Next, we seek travelling wave solutions of the form $g = \sqrt{\nu^2 - 1}\, G(\xi)$ where $\xi = (\nu^2-1)(\bar t-\bar z/\nu)$ involves the dimensionless wave velocity $\nu \geq 1$. Injecting this Ansatz in the above partial differential equation and integrating twice with respect to $\xi$ with vanishing integration constants yields a nonlinear differential equation for the strain:
\begin{equation}
	G = G^3 + \left(1 + \alpha G^2\right) \tfrac{\text d}{{\text d}\xi} G ,
	\label{KinkWave}
\end{equation}
where $\alpha = \frac32\delta(\nu^2-1)/\beta$ is a parameter.

From the above differential equation, one observes that travelling wave solutions to the wave equation \eqref{WaveStrain} should connect the equilibrium strains $G = 0$ and $G = \pm 1$ by following a smooth transition that depends on the parameter $\alpha$. Solutions read \citep{destrade13}
\begin{equation}
	\xi = -\ln\left[ \frac{1}{2G} \left(\frac43 (1-G^2)\right)^\frac{1+\alpha}2 \right]
	\label{SolWave}
\end{equation}
in implicit form, where we have enforced $G(0) = \frac12$ without loss of generality. Illustrations are provided later on.

\subsection{Slow time approximation}

\begin{figure}
	\centering
	\includegraphics{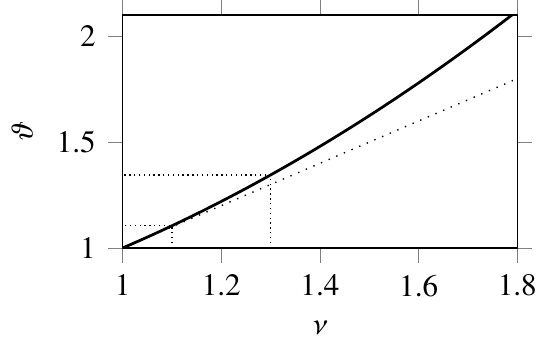}
	
	\caption{Scaled velocity $\vartheta = 1+\frac12 (\nu^2 - 1)$ for the `slow-time' reduced model in terms of the scaled velocity $\nu$ for the full wave equation. \label{fig:Speed}}
\end{figure}

In a similar fashion, let us now seek travelling wave solutions to the reduced wave equation \eqref{SlowStrain}. Thus, we first perform the substitutions \eqref{Scale} to get
\begin{equation}
	\frac{\partial g}{\partial \bar t} + \frac{\partial}{\partial \bar z} \left(g + \frac12 g^3\right) = \frac{1}{2} \frac{\partial^2 g}{\partial \bar z^2} .
	\label{SlowStrainScaled}
\end{equation}
In order to obtain wave solutions that correspond to the same strain values at infinity as in Sec.~\ref{subsec:KinkWave}, we introduce a slightly different scaling. Indeed, let us inject the Ansatz $g = \sqrt{\nu^2 - 1}\, G(\chi)$ with $\chi = (\nu^2-1) (\vartheta \bar t - \bar z)$ in Eq.~\eqref{SlowStrainScaled}, where $\vartheta = 1+\frac12 (\nu^2 - 1)$ is the new dimensionless velocity (Fig.~\ref{fig:Speed}). Thus, we arrive at the differential equation
\begin{equation}
	G = G^3 + \tfrac{\text d}{{\text d}\chi} G
	\label{KinkZabo}
\end{equation}
of which the strain values $0$ and $1$ are steady states. Enforcing the initial value $G = \frac12$ at $\chi = 0$ gives
\begin{equation}
	G = \frac{1}{\sqrt{1 + 3\, \text e^{-2 \chi}}} ,
	\label{SolZabo}
\end{equation}
which does not involve any extra parameter. One observes that this expression corresponds to the case $\alpha=0$ in Eqs.~\eqref{KinkWave}-\eqref{SolWave}.

\begin{remark}
	One might proceed in a similar fashion with the Burgers-type equation \eqref{Burgers} corresponding to the \emph{slow space} approximation. Similarly to \eqref{Scale}, we perform the substitutions $r(\bar z, \bar t) = \sqrt{2\beta/3}\, v(z,t)/c$ in Eq.~\eqref{Burgers} to get
	\begin{equation}
		\frac{\partial r}{\partial \bar z} + \frac{\partial}{\partial \bar t} \left(r - \frac12 r^3\right) = \frac{1}{2} \frac{\partial^2 r}{\partial \bar t^2} .
	\end{equation}
	Next, we introduce $r = \sqrt{\nu^2 - 1}\, V(\psi)$ where $\psi = (\nu^2-1) (\bar t - \bar z/\kappa)$ involves the dimensionless velocity $\kappa$ defined by the relationship $\kappa^{-1} = 1 - \frac12(\nu^2 - 1)$. This way, we obtain the same differential equation $V = V^3 + \frac{\text d}{{\text d}\psi} V$ for the dimensionless velocity $V$ as previously for the strain \eqref{KinkZabo}. Therefore, within the scope of the present study, the slow time and slow space approximations lead to related travelling wave solutions that describe the evolution of distinct kinematic variables (strain and velocity, respectively).
\end{remark}

\subsection{Comparison}\label{subsec:Compa}

Let us compare the solutions \eqref{SolWave}-\eqref{SolZabo} obtained for the full wave equation \eqref{WaveStrain} and the one-way model \eqref{SlowStrain}. First, one observes that these travelling waves of same amplitude do not propagate at the same speed, independently on the value of the parameter $\delta$ (Fig.~\ref{fig:Speed}). Indeed, given the expression of $\vartheta$, we can express the relative error $\mathscr{E} = \vartheta/\nu - 1$ on the scaled velocity as a function of $\nu$. To ensure that the latter remains less than 5\% (respectively 1\%), we obtain the requirement $\nu \leq 1.3$ (resp. $\nu \leq 1.1$) marked by dotted lines in the figure.

Now, let us observe that for a unit kink covering the range $0\leq G\leq 1$, the corresponding shear strains satisfy
\begin{equation}
	0 \leq \gamma \sqrt{\beta} \leq \gamma_{\max}\sqrt{\beta}, \qquad \gamma_{\max} = \sqrt{\alpha/\delta},
	\label{Range}
\end{equation}
where $\alpha = \frac32\delta(\nu^2-1)/\beta$ was introduced earlier on, see Eq.~\eqref{Scale}. In other words, in the case $\delta = 2$, the coefficient $\alpha$ in the differential equation \eqref{KinkWave} is related to the maximum strain $\gamma_{\max}$ of travelling waves, and these bounds are valid for both wave equations at hand due to application of the rescaling procedure \eqref{Scale}. Thus, restrictions of the wave speed $\nu$ can be expressed in terms of the strain. To ensure that the velocity error $\mathscr{E}$ remains less than 5\% (respectively 1\%), we therefore require $\gamma \sqrt{\beta} \leq 1.0$ (resp. $\gamma \sqrt{\beta} \leq 0.56$). Note that the parameter of nonlinearity can take such values as $\beta \approx 10$ for gels \citep{jacob07}. Therefore, the slow scale approximation has a very restricted validity for soft viscoelastic materials with strain-dependent shear viscosity ($\delta = 2$).

\begin{figure}
	\centering
	\includegraphics{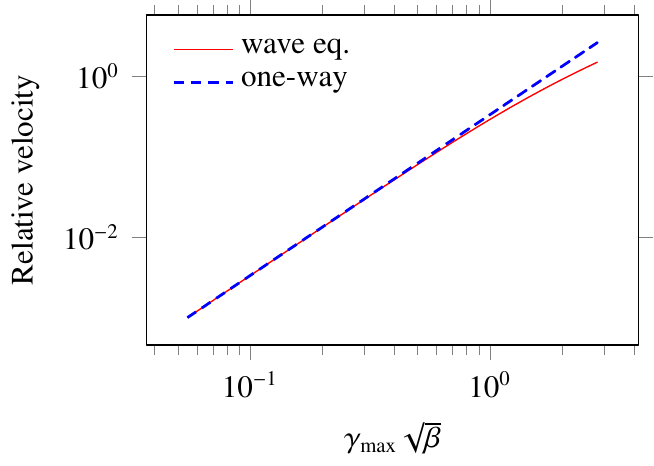}
	
	\caption{For the full wave equation with $\delta = 2$ (solid line) and the `slow-time' reduced model (dashed line), we represent the evolution of the relative velocity $\nu - 1$ (respectively, $\vartheta - 1$) of travelling waves in terms of the parameter $\gamma_{\max}\sqrt{\beta}$ which involves the strain amplitude and the coefficient of nonlinearity. The axes have a logarithmic scale. \label{fig:Evol}}
\end{figure}

This property is further illustrated in Fig.~\ref{fig:Evol}, where we have represented the evolution of the relative velocity $\nu - 1$ (or $\vartheta - 1$) in terms of the maximum strain amplitude, both for the full wave equation with $\delta=2$ and its one-way approximation. According to the expression of $\gamma_{\max}$ above, we have the relationship $\vartheta - 1 = \frac13 (\gamma_{\max}\sqrt{\beta})^2$ in the case of the one-way approximate model, which produces lines of slope two in log-log coordinates (dashed lines in the figure). However, for the full wave equation with $\delta=2$, this relationship between the wave speed $\nu$ and the strain amplitude is not satisfied. Differences between the one-way model and the full wave equation become visible at large strains.

\begin{figure}
	\centering
	\includegraphics{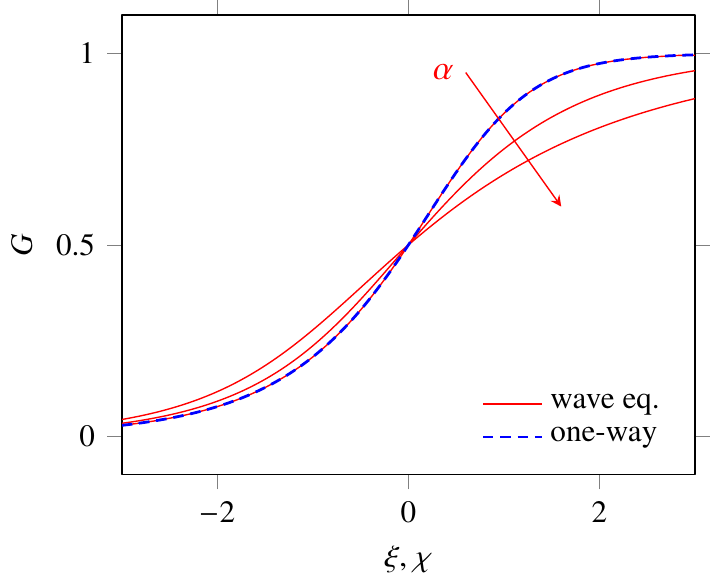}
	
	\caption{Steady waveforms deduced from Eqs.~\eqref{SolWave}-\eqref{SolZabo} for increasing values of the parameter $0\leq \alpha \leq 3$ (arrow). Evolution of the scaled shear strain $G$ in terms of the related dimensionless coordinate $\xi$ or $\chi$. \label{fig:Comp}}
\end{figure}

In Fig.~\ref{fig:Comp}, we display the evolution of the waveforms \eqref{SolWave}-\eqref{SolZabo} in terms of the scaled coordinates $\xi$, $\chi$. In the case of the full wave equation \eqref{KinkWave}, the parameter $\alpha$ takes the values $\lbrace 0, 1.2, 3 \rbrace$. It appears that the waveforms so-obtained follow a drastically different evolution when parameters are modified. In particular, the wavefront deduced from the full wave equation (solid lines) does not necessarily exhibit the same invariance and symmetry properties as the wavefront deduced from the one-way model (dashed line).

\section{Inviscid limit}\label{sec:Elast}

\subsection{Simple waves}\label{subsec:Simple}

In the lossless case, exact one-way wave equations can be derived by using the method of Riemann invariants, see for instance the introductory example by \citet{john76}. Such particular wave solutions called \emph{simple waves} keep one Riemann invariant constant. In other words, the particle velocity $v = R - Q(\gamma)$ with $Q(\gamma) = c \int_0^\gamma \sqrt{1+2\beta g^2}\, \text d g$ depends explicitly on the strain $\gamma$. The scalar $R$ is an arbitrary constant, for instance $R \equiv 0$ in some specific boundary-value problems \citep{berjamin22}, which will be assumed satisfied from now on. Spatial differentiation of the velocity then produces
\begin{equation}
	\frac{\partial\gamma}{\partial t} + c \sqrt{1+2\beta\gamma^2}\, \frac{\partial \gamma}{\partial z} = 0,
	\label{Riemann}
\end{equation}
where we have used the equality of mixed partials $\partial v/\partial z = \partial \gamma/\partial t$. Obviously, the lossless one-way wave equation \eqref{SlowStrain} with $\tau = 0$ is an approximation of \eqref{Riemann} for $2\beta\gamma^2 \ll 1$.

Let us analyse this requirement in a more quantitative manner. To ensure that the relative error on the advection velocity $\mathscr{E} = \frac{1+a}{\sqrt{1+2a}} - 1$ for $a=\beta\gamma^2$ remains less than 5\% (respectively 1\%), we obtain the requirement $a \leq 0.44$ (resp. $a \leq 0.16$). Application of the square root leads to the restriction $\gamma\sqrt{\beta} \leq 0.66$ (resp. $\gamma\sqrt{\beta} \leq 0.40$) which is slightly more constraining than in the case of viscoelastic travelling waves (Sec.~\ref{subsec:Compa}).

Along a simple wave, computation of the partial derivative of the velocity $v = -Q(\gamma)$ with respect to time produces
\begin{equation}
	c \frac{\partial v}{\partial z} + \left(1+2\beta\gamma^2\right)^{-1/2} \frac{\partial v}{\partial t} = 0,
	\label{RiemannV}
\end{equation}
where the strain $\gamma = Q^{-1}(-v)$ can be expressed formally as a function of the velocity, despite no analytical expression of the inverse function $Q^{-1}$ of $Q$ is known in the present case. If $|v|$ is small, then we can use the approximation $\gamma \simeq -v/c$ of the strain which follows from the asymptotic equivalence of $Q \sim c\gamma$ at small strains. Next, the $(\cdot)^{-1/2}$-factor in Eq.~\eqref{RiemannV} can be approximated by the polynomial expression $1-\beta\gamma^2$ as long as $2\beta \gamma^2 \ll 1$. This way, we have shown that the one-way wave equation \eqref{Burgers} is an approximation of Eq.~\eqref{RiemannV} obtained for $R = 0$ and $2\beta v^2/c^2 \ll 1$ in the elastic limit. This observation is consistent with the discussions in \citet{catheline03e}. In summary, the lossless `slow-space' and `slow-time' reductions \eqref{Burgers}-\eqref{SlowStrain} with $\tau = 0$ are approximate governing equations for simple waves with small values of $\beta v^2/c^2$ and of $\beta \gamma^2$, respectively.

The general solution of the scalar wave equations \eqref{SlowStrain} and \eqref{Riemann} can be obtained by using the method of characteristics. For instance, let us consider an initial-value problem, and assume that the initial strain is a triangular bump of width $L$ and of amplitude $\gamma_{\max} = K/\sqrt{2\beta}$ where $K$ is a parameter. In other words, $\gamma(z,0)$ equals $\gamma_{\max} \big(1 - 2\frac{|z|}{L}\big)$ for $|z|\leq L/2$, and it equals zero elsewhere.

Let us first solve this problem for the simple wave governed by Eq.~\eqref{Riemann}. Introducing the dimensionless coordinates $\breve{z} = 2z/L$ and $\breve{t} = 2ct/L$, the strain solution so-obtained reads
\begin{equation}
	\frac{\gamma(z,t)}{\gamma_{\max}} =
	\left\lbrace\begin{aligned}
		& \frac{1\mp \breve{z}\pm \breve{t}\sqrt{1 + K^2 \left((1\mp \breve{z})^2-\breve{t}^2\right)}}{1-K^2\breve{t}^2},\\
		&\text{or}\quad \frac{(1\mp \breve{z})^2-\breve{t}^2}{2\, (1\mp \breve{z})} \quad\text{if}\quad K^2\breve{t}^2= 1,
	\end{aligned}\right.
	\label{RiemannSol}
\end{equation}
for $|\breve{z}-\breve{t}| \leq 1$, zero elsewhere, where the expressions with the top sign (respectively, the bottom sign) are defined over the domain $\breve{z} - \breve{t}\sqrt{1+K^2} \geq 0$ (resp. $\leq 0$). This solution is only valid for small times $\breve{t} \leq \breve{t}^\star$ beyond which it becomes multi-valued. The critical time $\breve{t}^\star = \frac{1 + \sqrt{1+K^2}}{K^2}$ is the shock formation time.

For the slow time approximation \eqref{SlowStrain} with $\tau = 0$, the solution of the initial-value problem reads
\begin{equation}
	\frac{\gamma(z,t)}{\gamma_{\max}} = \frac{1 - \sqrt{1 + 2K^2 \breve{t}\, (\breve{z}-\breve{t} \mp 1)}}{\pm K^2\breve{t}} ,
	\label{SlowSol}
\end{equation}
for $|\breve{z}-\breve{t}| \leq 1$, zero elsewhere, where the expression with the top sign (resp. the bottom sign) is defined over the domain $\breve{z} - \breve{t} (1+\tfrac12 K^2) \geq 0$ (resp. $\leq 0$). This solution is valid for small times $\breve{t} \leq 2/K^2$ at which a shock wave is formed.

\begin{figure}
	\centering
	\includegraphics{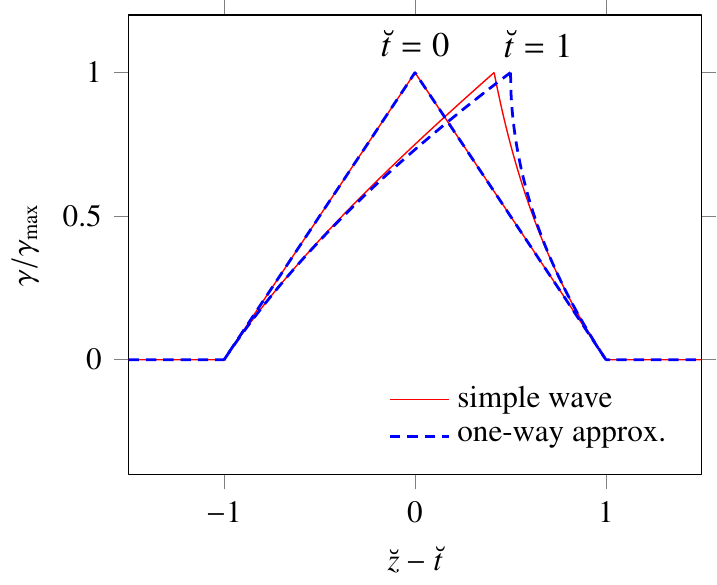}
	
	\caption{Inviscid case. Evolution of the waveforms from initial triangular strain with increasing dimensionless times $\breve{t}$. \label{fig:CompIVP}}
\end{figure}

Fig.~\ref{fig:CompIVP} displays the solutions \eqref{RiemannSol}-\eqref{SlowSol} at increasing times $\breve{t}$ where we have set $K=1$. Comparison of the waveforms suggests that the error increases pointwise with time. This observation can be expressed in a more quantitative fashion. In fact, a Taylor series expansion of the difference $\eqref{SlowSol}-\eqref{RiemannSol}$ with respect to the strain amplitude parameter $K = \gamma_{\max}\sqrt{2\beta}$ shows that the slow time solution \eqref{SlowSol} approximates the simple wave \eqref{RiemannSol} correctly up to a term of order $O(K^4)$, pointwise. The leading-order coefficient in this expansion is proportional to $\breve{t}\, |\breve{z}-\breve{t}\pm 1|^4$, which shows that the error increases with time, and as one moves away from the domain where the solution is equal to zero (cf. Figure). Possibly, computation of the global error by spatial summation of the pointwise errors would be insightful.

\subsection{Shock waves}\label{subsec:Shock}

The process leading to simple wave solutions requires some smoothness of the waveform, an assumption that will be relaxed in this subsection. Indeed, based on the equations of motion, the speed $\sigma$ of a discontinuity must satisfy the Rankine--Hugoniot jump conditions
\begin{equation}
	-\llbracket v\rrbracket = \sigma \llbracket \gamma\rrbracket ,\qquad
	-\llbracket P^\text{e}_{13}\rrbracket = \rho\sigma \llbracket v \rrbracket ,
\end{equation}
where $\llbracket\cdot\rrbracket = (\cdot)^+ - (\cdot)^-$ denotes the jump of a quantity across the discontinuity. Thus, $\rho \sigma^2 \llbracket \gamma\rrbracket = \llbracket P^\text{e}_{13}\rrbracket$. For a right-going discontinuity $\sigma \geq 0$ propagating into an undeformed domain $(\gamma)^+ = 0$, we find
\begin{equation}
	\sigma = c \sqrt{1 + \tfrac23\beta  (\gamma^-)^2}  ,
	\label{WaveRH}
\end{equation}
where \eqref{WaveCoeffs} and the expression of $P^\text{e}_{13}$ in Sec.~\ref{sec:Shear} were used. Application of Liu's entropy criterion shows that such a shock wave is admissible \citep{berjamin17}.

Let us now consider the one-way approximate wave equation \eqref{SlowStrain} with $\tau = 0$, which can be rewritten in conservation form $\partial_t\gamma = -\partial_z f(\gamma)$ for a suitable definition of the flux $f$. In this case, the Rankine--Hugoniot condition for a discontinuity travelling at speed $\sigma = \llbracket f\rrbracket / \llbracket \gamma\rrbracket$ yields
\begin{equation}
	\sigma = c \left(1 + \tfrac13\beta (\gamma^-)^2 \right) .
	\label{BurgersRH}
\end{equation}
Obviously, Eq.~\eqref{BurgersRH} is an approximation of Eq.~\eqref{WaveRH} for $\tfrac23\beta (\gamma^-)^2 \ll 1$. By setting $a = \frac13\beta(\gamma^-)^2$, the expression of the relative wave velocity error $\mathscr{E}$ from Sec.~\ref{subsec:Simple} still applies. Furthermore, the study of initial-value problems with discontinuous solutions might be conducted in a similar fashion as in the previous section.


\section{Conclusion}\label{sec:Conclu}

For a specific strain-rate viscoelasticity theory of soft solids, we have shown that one-way approximate wave propagation models can produce significantly different travelling wave solutions than the full equations of motion as soon as the wave amplitude is not infinitesimal. Similar observations are reported in the literature in relation with shear shock formation \citep{berjamin22}. In the elastic limit, we have examined the validity of one-way approximations in relation with simple wave theory and shock wave theory, thus leading to dedicated criteria involving small velocity and strain amplitudes. We conclude that these approximations should be used with care given their limited accuracy, in general. Nevertheless, they might remain useful for the interpretation of experimental results where their validity is not always severely penalised \citep{catheline03,catheline03e}.

Due to the assumption of slowly-varying wave profiles, Burgers-type model equations have a limited ability to account for dispersion effects, which play an important role in viscoelasticity (Fig.~\ref{fig:Dispersion}). Besides viscous stresses, other dispersive effects might be impacted by the rescaling procedure. For instance, one might insert the dispersive term $\hat\tau^2\, \partial^4 \gamma/\partial z^2\partial t^2$ in the right-hand side of the wave equation \eqref{WaveStrain}, see \citet{destrade08} for related theories. With $\hat\tau^2$ of order $\epsilon^2$, this assumption leads to the addition of $\frac12 \hat\tau^2\, \partial^3 v/\partial t^3$ in the right-hand side of \eqref{Burgers}, and to the addition of $-\frac12 c^3\hat\tau^2\, \partial^3 \gamma/\partial z^3$ in the right-hand side of \eqref{SlowStrain}. Consistently, dispersion analysis shows that the slow scale equations \eqref{Burgers}-\eqref{SlowStrain} with dispersion are valid at low frequency only.

Within the present modelling framework, more general constitutive theories could be considered, e.g. to account for material anisotropy \citep{destrade10b}. Moreover, one could consider a generalised model that accounts for stress relaxation \citep{saccomandi21}. Lastly, a similar discussion could be proposed for the partial differential equations governing the diffraction of unidirectional wave beams \citep{wochner08}, see also the case of the KZK equation \citep{rozanova09}.

\section*{Acknowledgments}
	The author is grateful to Michel Destrade (Galway, Ireland) for support, and to the anonymous reviewers for their valuable suggestions. This project has received funding from the European Union's Horizon 2020 research and innovation programme under grant agreement TBI-WAVES --- H2020-MSCA-IF-2020 project No. 101023950.



\appendix
\section{Consequences of incompressibility}\label{app:Incomp}

This paragraph is devoted to the derivation of Eq.~\eqref{InvarViscoIncomp}. We start with the Cayley--Hamilton identity for the right Cauchy--Green tensor $\bm{C} = \bm{F}^\transpose\bm{F}$, which reads
\begin{equation}
	\bm{C}^3 - \textsl{I } \bm{C}^2 + \textsl{II } \bm{C} - \textsl{III } \bm{I} = \bm{0}, 
	\label{CH}
\end{equation}
where $\textsl{I}$, $\textsl{II}$, $\textsl{III}$ are the principal invariants of $\bm{C}$. In the case of volume-preserving motions \eqref{J}, the tensor $\bm C$ is unimodular, i.e. we have $\textsl{III} = 1$. Next, multiplication of \eqref{CH} by $\bm{C}^{-1}\dot{\bm E}$ on the right side and substitution of $\bm{C} = \bm{I}+2\bm{E}$ lead to
\begin{equation}
	\bm{C}^{-1}\dot{\bm E} = (\textsl{II}- \textsl{I}+1) \dot{\bm E} + (4-2\textsl{I})\bm{E}\dot{\bm E} + 4\bm{E}^2 \dot{\bm E} .
	\label{Calc}
\end{equation}
Computation of the trace entails \eqref{InvarViscoIncomp}, where we have used the incompressibility property $\text{tr}\, \bm{D} = \text{tr}(\bm{C}^{-1}\dot{\bm E}) = 0$, the definition of the invariants \eqref{InvarElast}-\eqref{InvarVisco} and the relationship between $\textsl{I}$, $\textsl{II}$ and the invariants $I_k$ used here \citep{destrade10c}.

Now, let us briefly discuss the derivation of the Newtonian viscous stress \eqref{Fluid}. For this purpose, we introduce the strain tensor $\bm{\Pi} = \bm{C}^{-1}\bm{E}$, which satisfies
\begin{equation}
	\begin{aligned}
		&\dot{\bm \Pi} = \bm{C}^{-1}\dot{\bm E}\bm{C}^{-1} = \bm{F}^{-1} \bm{D} \bm{F}^{-\text{T}} ,\\
		&\dot{\bm \Pi}\dot{\bm E} = (\bm{C}^{-1}\dot{\bm E})^2 = \bm{F}^{-1} \bm{D}^2 \bm{F} .
	\end{aligned}
\end{equation}
These identities lead to the expression \eqref{Fluid} of the viscous stress and of the dissipation potential $W^\text{v} = \eta \text{ tr}(\dot{\bm \Pi}\dot{\bm E})$. Squaring \eqref{Calc} and computing the trace yields a lengthy expression for $W^\text{v}$ (hint: write the Cayley--Hamilton identity for $\bm{E}$), which satisfies $W^\text{v} \simeq \eta I_7$ at leading order in $\bm{E}$, $\dot{\bm E}$. However, the Newtonian theory \eqref{Fluid} and the viscous stress \eqref{Stresses}\textsubscript{b} might differ at higher order.


\bibliographystyle{plainnat}
\bibliography{Biblio}{}

\end{document}